# Thermodynamics of Climate Change: Generalized Sensitivities

Valerio Lucarini [v.lucarini@reading.ac.uk][1,2,3], Klaus Fraedrich[4] & Frank Lunkeit[4]

[1]Department of Meteorology, University of Reading, Earley Gate, PO Box 243, Reading RG6 6BB (UK); [2]Department of Mathematics, University of Reading, Whiteknights, PO Box 220 Earley Gate, PO BOX 243, Reading RG6 6AX (UK); [3]Department of Physics, University of Bologna, Viale Berti Pichat 6/2, 40127 Bologna (Italy); [4] Meteorologisches Institut, Klima Campus, University of Hamburg, Grindelberg 5, 20144 Hamburg (Germany)

Using a recently developed formalism, we present an in-depth analysis of how the thermodynamics of the climate system varies with $CO_2$ concentration by performing experiments with a simplified yet Earth-like climate model. We find that, in addition to the globally averaged surface temperature, the intensity of the Lorenz energy cycle, the Carnot efficiency, the entropy production and the degree of irreversibility of the system are linear with the logarithm of the $CO_2$ concentration. The generalized sensitivities proposed here suggest that the climate system becomes less efficient, more irreversible, and features higher entropy production as it becomes warmer.

## Introduction

The most basic way to characterize the climate system is describing it as a non-equilibrium thermodynamic system, generating entropy by irreversible processes and - if time-dependent forcings can be neglected - keeping a steady state by balancing the input and output of energy and entropy with the surrounding environment.

A primary goal of climate science is to understand how the statistical properties of the climate system change as a result of variations in the value of external or internal parameters. Rigorous mathematical foundations to this problem can be traced to the recent introduction of the Ruelle response theory[22] for non equilibrium steady state systems[23]. Such a response theory has been recently proved to have formal analogies with the usual Kubo response theory for quasi-equilibrium systems[24] and to be amenable to numerical investigation[25].

Correspondingly, it has long been recognized that a prominent way of establishing a comprehensive view on the climate system relies on adopting a thermodynamic perspective. Two main approaches can be envisioned along this line.

In the first approach, the focus is on the dynamical mechanisms and physical processes responsible for the transformation of energy from one form to the other. The concept of the energy cycle of the atmosphere introduced by Lorenz[1] allowed for defining an effective climate machine, driven by the temperature difference between a warm and a cold thermal pool, such that the atmospheric and oceanic motions are at the same time the result of the mechanical work (then dissipated by in a turbulent cascade) produced by the engine, and are processes which re-equilibrate the energy balance of the climate system[2]. More recently, advances along this line have come from Winters et al[3] who dealt more specifically with ocean-like conditions, and, especially, from Johnson[4], who introduced a rather convincing Carnot engine–equivalent picture of the climate system by defining the warm and the cold reservoirs and their temperatures on a physically sound and computationally feasible basis.

In the second approach, the emphasis lies on the analysis of the irreversibility of the climate system, and, especially, of its entropy production. This largely results from the intellectual stimulation coming from the maximum entropy production principle (MEPP), first discussed by Paltridge[5], which proposes that an out-of-equilibrium nonlinear system adjusts in such a way to maximize the production of entropy[6]. stimulations is based upon the non-equilibrium thermodynamics to the climate system. Even if the validity of MEPP – in general and in the climate system in particular - is unclear[7,8], its heuristic adoption has been quite fruitful[9,10,11,12] and, more importantly, has stimulated a detailed re-examination of the importance of entropy production in the climate system and an active effort towards giving quantitative estimates[2,13]. Moreover, this has resulted into a drive for adopting of a new generation of diagnostic tools based on the 2$^{nd}$ law of thermodynamics for auditing climate models[14,15] and outline a set of parameterizations to be used in conceptual and intermediate complexity models or for the reconstruction of the past climate conditions[16].

Recently, using tools borrowed from classical thermodynamics, a link has been found between the Carnot efficiency, the entropy production and the degree of irreversibility of the climate system[17]. This has made possible a new exploration[18] of the onset and decay of snowball conditions[18] as parametrically controlled by variations in the solar constant using a simplified and portable climate model[20,21]. In that analysis, the two branches of cold and warm climate stationary states have been found to feature very distinct macro-thermodynamical properties and the tools proposed[17] have proved to allow for a relevant insight into this paleoclimatic problem.

In an attempt to make a synthesis of various research lines, in this paper we revise and revive the classic problem of analyzing the climate sensitivity to $CO_2$ concentration changes by adding on top of the usual, IPCC-like[26] analysis of globally averaged surface temperature changes the investigation of how the global thermodynamics of the system is influenced by the atmospheric composition, so that a wider and better physically-based set of generalized sensitivities are introduced. Our investigation is performed, along the lines of[18], using the simplified and portable climate model PLASIM[20,21]. We believe our work contributes to presenting reliable metrics to be used in the validation of climate models of various degrees of complexity[14,15].

## Efficiency and Entropy Production in the Climate System

Following Johnson[4], we define the total energy of the $\Omega$-domain of the climatic system by $E(\Omega)=P(\Omega)+K(\Omega)$, where *P* represents the moist static potential energy, given by the thermal – inclusive of the contributions due to water phase transitions – and potential contributions, and *K* is the total kinetic energy. The time derivative of the total kinetic and potential energy can be expressed as $\dot{K}=-\dot{D}+\dot{W}, \dot{P}=\dot{\Psi}+\dot{D}-\dot{W}$, where we have dropped $\Omega$-dependence for convenience, $\dot{D}$ is the (positive definite) integrated dissipation, $\dot{W}$ is the instantaneous work performed by the system (or, in other term, the total intensity of the Lorenz energy cycle[1]), $\dot{\Psi}$ represents the hearting due to convergence of heat fluxes (which can be split into the radiative, sensible, and latent heat components), such that $\dot{E}=\dot{\Psi}$. We denote the total heating rate as $\dot{\Phi}=\dot{\Psi}+\dot{D}$. Under the hypothesis of a non-equilibrium steady state system[23], we have $\overline{\dot{E}}=\overline{\dot{P}}=\overline{\dot{K}}=0$, where the upper bar indicates time averaging over a long time scale. At any instant, we can partition the domain $\Omega$ into $\Omega^{+}$ and $\Omega^{-}$, such that the intensive total heating rate $\dot{Q}$ is positive in $\Omega^{+}$ and negative in $\Omega^{-}$:

$$\dot{P}+\dot{W}=\int_{\Omega^{+}} dV\rho\dot{Q}^{+}+\int_{\Omega^{-}} dV\rho\dot{Q}^{-}=\dot{\Phi}^{+}+\dot{\Phi}^{-}=\dot{\Phi}. \qquad (1)$$

Since dissipation is positive definite, we obtain $\overline{\dot{W}} = \overline{\dot{D}} = \overline{\dot{\Phi}^+} + \overline{\dot{\Phi}^-} > 0$. Assuming local thermodynamic equilibrium – which applies pretty well everywhere except in the upper atmosphere, which has a negligible mass – and, neglecting the impact of mixing processes, which seems negligible on the global scale[17], we have that locally $\dot{Q} = \dot{s}T$, so that at any instant entropy fluctuations have locally the same sign as heat fluctuations. The derivative of the entropy of the system can be written as $\dot{S} = \dot{\Sigma}^+ + \dot{\Sigma}^-$, where the signs indicate the positive and negative definite integral contributions. At steady state $\overline{\dot{S}} = 0$, so that $\overline{\dot{\Sigma}^+} = -\overline{\dot{\Sigma}^-}$, with $\overline{\dot{\Sigma}^+}$ measuring the absolute value of the entropy fluctuations throughout the domain. The mean value theorem allows for expressing $\overline{\dot{\Sigma}^+} = \overline{\dot{\Phi}^+}/\Theta^+$ and $\overline{\dot{\Sigma}^-} = \overline{\dot{\Phi}^-}/\Theta^-$, where $\Theta^+$ ($\Theta^-$) is the time and space averaged value of the temperature where absorption (release) of heat occurs. Since $\left|\overline{\dot{\Sigma}^+}\right| = \left|\overline{\dot{\Sigma}^-}\right|$ and $\left|\overline{\dot{\Phi}^+}\right| > \left|\overline{\dot{\Phi}^-}\right|$, we derive that $\Theta^+ > \Theta^-$ and we obtain that the Lorenz energy cycle can be reformulated as the result of a Carnot engine such that $\overline{\dot{W}} = \eta\overline{\dot{\Phi}^+}$, where $\eta = \left(\Theta^+ - \Theta^-\right)/\Theta^+ = \left(\overline{\dot{\Phi}^+} + \overline{\dot{\Phi}^-}\right)/\overline{\dot{\Phi}^+}$ is the efficiency[4,17].

The 2nd law of thermodynamics imposes that, when considering the long-term average of the material entropy production inside the system[13] $\overline{\dot{S}_{in}(\Omega)}$, its minimal value can be estimated[17] $\overline{\dot{S}_{\min}(\Omega)} \approx \overline{\dot{W}}/\langle\Theta\rangle \approx \eta\overline{\dot{\Sigma}^+}$, where $\langle\Theta\rangle = \left(\Theta^+ + \Theta^-\right)/2$, and linked to the viscous dissipation of kinetic energy. Therefore, $\eta$ sets also the scale relating the minimal entropy production of the system to the absolute value of the entropy fluctuations inside the system. If the system is isothermal and at equilibrium the internal entropy production is zero, since the efficiency $\eta$ is vanishing: the system has already attained the maximum entropy state. The excess of entropy production with respect to the minimum, $\overline{\dot{S}_{exc}}$, is due to the heat transport downgradient the temperature field. Therefore, we can define:

$$\alpha = \overline{\dot{S}_{exc}} \Big/ \overline{\dot{S}_{\min}} \approx \int_{\Omega} dV \overline{\vec{H} \cdot \vec{\nabla}\left(\frac{1}{T}\right)} \Big/ \left(\overline{\dot{W}} \big/ \langle \Theta \rangle\right) \approx \int_{\Omega} dV \overline{\vec{H} \cdot \vec{\nabla}\left(\frac{1}{T}\right)} \Big/ \left(\eta \overline{\dot{\Sigma}^{+}}\right) \geq 0 \qquad (2)$$

as a parameter of irreversibility of the system, which vanishes if all the production of entropy is due to the – unavoidable - viscous dissipation of the mechanical energy. Therefore, the more efficiently the system transports heat from high to low temperature regions, the larger is the entropy production, *ceteris paribus*. As $\overline{\dot{S}_{in}} \approx \eta \overline{\Sigma}^{+}(1+\alpha)$, we have that entropy production is maximized if we have a joint optimization of heat transport and of production of mechanical work. This formula provides a new way of interpreting the controversial MEPP[17].

## Results

In the usual operative definition, “climate sensitivity” $\Lambda_{T_S}$, is the increase of the globally averaged mean surface temperature $\overline{T_S}$ between the preindustrial $CO_2$ concentration steady state and the steady state conditions realized when $CO_2$ concentration is doubled[26]. As $\overline{T_S}$ is almost linear with respect to the logarithm of the $CO_2$ concentration[27] on a relatively large range, it is actually easy to generalize the definition of $\Lambda_{T_S}$ as the impact on $\overline{T_S}$ of $CO_2$ doubling, regardless of the initial concentration, so that $\Lambda_{T_S} = d\overline{T_S} \big/ d \log_2\left([CO_2]_{ppm}\right)$.

The main goal of this work is to check whether it is possible to define, in a similar fashion, generalized sensitivities $\Lambda_X$ to describe the steady state response to $CO_2$ concentration changes of the thermodynamical properties *X* of the climate system.

At this scope, we have performed climate simulations with PLASIM for $CO_2$ concentrations ranging from 50 to 1850 ppm by 50 ppm steps, thus totalling 37 runs, each lasting 50 years. We anticipate that no bistability regions – parametric ranges where the system features an attractor which can be split into two connected parts – have been found when altering the atmospheric composition, as opposed to a previous analysis where the solar constant has been varied within a paleoclimatically feasible range[18]. In order to fully characterize the non-equilibrium properties of the climate system,  we have analysed the most important thermodynamic variables of the system introduced in the previous section:

- the time average of the temperatures of the warm - $\Theta_1$ - and cold reservoir - $\Theta_2$, and, as a reference, the time average of the global mean surface temperature $\overline{T_S}$ (Fig. 1);
- the thermodynamic efficiency $\eta$ (Fig. 2a);
- the average intensity of the Lorenz energy cycle $\overline{\dot{W}}$ (Fig. 2b);
- the time average of material entropy production $\overline{\dot{S}_{in}}$ (Fig. 2c);
- the degree of irreversibility of the system $\alpha$ (Fig. 2d).

See the Methods section for additional details in the calculations. It is rather interesting to observe that, in addition to the surface temperature, all of these thermodynamic variables feature a striking linear behaviour with respect to the logarithm of the $CO_2$ concentration. Therefore, we can safely attribute a robust value (with an uncertainty of about 5-10%) to the generalized sensitivities defined as $\Lambda_X \equiv dX/d\log_2\left([CO_2]_{ppm}\right)$. Results are summarized in Table 1.

The three temperature indicators (Fig. 1) feature, as expected, positive sensitivities: the surface temperature sensitivity is well within the range of what simulated by the climate models included in IPCC[26], whereas the two bulk thermodynamic temperatures have smaller sensitivities. Therefore, an increase of the vertical temperature gradient is predicted for higher $CO_2$ concentrations. Moreover, as the temperature of the cold bath increases faster than that of the warm bath, higher $CO_2$ concentration implies a more isothermal atmosphere, in agreement with previous findings [18]. The main process contributing to this effect is large enhancement of latent heat fluxes due to the impact of increasing average temperature on the Clausius-Clapeyron relation. Consequently, increases in the $CO_2$ concentration cause a steep decrease in the efficiency of the climate system, as shown in Fig 2a. In the explored range, the efficiency decreases by about 35%, with a relative change of about –7% per $CO_2$ doubling. In a thicker (and warmer) atmosphere, the absorbed heat $\overline{\dot{\Phi}^{+}}$ is larger, so that the actual strength of the Lorenz energy cycle changes as the result of the competing effects of increasing energy input and decreasing efficiency. The intensity of the Lorenz cycle decreases in a warmer climate (Fig. 2b), with an approximate change of –4% per $CO_2$ doubling. By energy conservation, the same applies to the total dissipation, so that we in a warmer climate weaker surface winds are expected.

As with increasing $CO_2$ concentration the average temperature increases and the total dissipation decreases, $\overline{\dot{S}_{\min}}$ – which is related uniquely to mechanical dissipation – is a decreasing function of $CO_2$ concentration. Instead, as shown in Fig. 2c, the actual average rate of entropy production $\overline{\dot{S}_{in}}$ has the opposite behaviour, with an approximate relative increase of 2% per $CO_2$ doubling. This implies that the entropy production due to the heat transport downgradient the temperature field is much higher in warmer climates, the reason being, again, that latent heat fluxes become extremely effective in transporting heat. Therefore, the degree of irreversibility of the system $\alpha$ increases steeply with $CO_2$ concentration (Fig. 2d). In the considered range, the fraction of entropy production due to mechanical energy dissipation $1/(\alpha+1)$ drops from about 22% to about 12%. Note that this behaviour is specific for climate conditions analogous to the present ones, whereas under snowball conditions, where latent heat fluxes are negligible, higher temperatures lead to higher total entropy production, the value of $\alpha$ is about unity and only slightly affected by temperature[18].

## Conclusions

We have proposed a new approach for analyzing the classical problem of the steady-state response of the climate system to $CO_2$ concentration changes and have demonstrated its validity with a simplified yet Earth-like climate model. We have introduced a comprehensive set of generalized climate sensitivities describing the response of the global thermodynamical properties of the climate system, building upon a recently introduced solid theoretical framework[17].

We find that, in addition to the globally averaged surface temperature, the intensity of the Lorenz energy cycle, the Carnot efficiency, the entropy production and the degree of irreversibility of the system are linear with the logarithm of the $CO_2$ concentration. The generalized sensitivities proposed here (whose values are reported in Table 1) demonstrate that the climate system becomes less efficient, more irreversible, and features higher entropy production as it becomes warmer. Changes in intensity of the latent heat fluxes tend to be the dominating ingredients, thus showing, at a fundamental level, how important it is to address correctly the impact of climate change on the hydrological cycle.

Due to the monotonic (and, in particular, linear) dependence of the diagnosed variables with respect to the logarithm of the $CO_2$ concentration, it is possible to re-parameterize efficiently all the variables with respect to just this one. In Table 2, we also provide the linear coefficients of all the thermodynamic macro-variables of the system with respect to the changing surface temperature. These data may be of use when devising simplified yet comprehensive climate models or estimating unknown quantities in comprehensive models or from actual observational data.

We believe that the investigation proposed here may serve as a stimulation to re-examine, from a more fundamental point of view, the problem of climate change, and may help addressing problems of paleoclimatological relevance, such as the interplay between solar constant and atmospheric composition changes in determing ice ages, or the onset and the decay of snowball conditions. We expect that the extensive application of the thermodynamically-based tools adopted here may, in general, help closing the "Gap between Simulation and Understanding in Climate Modeling"[14] and may constitute the basis for a new generation of metrics aimed at the validation of climate models[15].

## Appendix - Methods

The Planet Simulator[20,21] (http://www.mi.uni-hamburg.de/plasim) is a climate model of intermediate complexity. The primitive equations formulated for vorticity, divergence, temperature and the logarithm of surface pressure are solved via the spectral transform method. The mode is used in a configuration featuring T21 horizontal resolution with five sigma levels in the vertical. The parameterizations for unresolved processes consist of long and short wave radiation, interactive clouds, moist and dry convection, large scale precipitation, boundary layer fluxes of latent and sensible heat and vertical and horizontal diffusion. The land surface scheme uses five diffusive layers for the temperature and a bucket model for the soil hydrology. The ocean is represented by a 50m mixed-layer (swamp) ocean (with energy transport set to 0) which includes a 0-dimensional thermodynamic sea ice model. As recently discussed by Danabasoglu and Gent[28], slab ocean climate models are well suited for providing an accurate steady state climate response. Following Lucarini and Fraedrich[29], the global atmospheric energy balance is greatly improved with respect to previous versions of the model by re-feeding the kinetic energy losses due to surface friction and horizontal and vertical momentum diffusion. See Becker for an in-depth discussion on these topics[30]. PLASIM performs satisfactorily well, as the average energy bias is in all simulations smaller than $0.2\,Wm^{-2}$, which is about one order of magnitude smaller than most state-of-the art climate models[31]. This is done locally by an instantaneous heating of the environmental air. Each simulation has a length of 50 years to ensure the system to archive equilibrium well before the end of the run. The statistical properties of the observables considered in this study are computed on the last 30 years of the simulations in order to rule out the presence of transient effects. Simulations are performed with $CO_2$ concentration set to 50 to 1850 ppm with steps of 50 ppm. We consider, in addition to the global, time averaged surface temperature $\overline{T_s}$ the following bulk thermodynamic observables: $\overline{\dot{\Sigma}^+}$, $\overline{\dot{\Sigma}^-}$, $\overline{\dot{\Phi}^-}$, $\overline{\dot{\Phi}^+}$ and $\overline{\dot{S}_{in}}$ .

In spite to the improvements to the energy budget described above, PLASIM presents some modest biases in the global energy and entropy budgets, so that $\overline{\dot{S}} = \overline{\dot{\Sigma}^+} + \overline{\dot{\Sigma}^-} = \Delta_S$, with $\Delta_S << \left|\overline{\dot{\Sigma}^+}\right|, \left|\overline{\dot{\Sigma}^-}\right|$, and that $\overline{\dot{E}} = \overline{\dot{P}} + \overline{\dot{K}} = \Delta_E$, with $\Delta_E << \left|\overline{\dot{\Phi}^+}\right|, \left|\overline{\dot{\Phi}^-}\right|$, so that the NESS hypothesis is not obeyed. Due to these biases, the

Lorenz energy cycle has a spurious term source/sink term, so that $\overline{\dot{W}} = \overline{\dot{\Phi}^{+}} + \overline{\dot{\Phi}^{-}} - \Delta_E$. Therefore, the thermodynamic efficiency is ill defined and, similarly, the estimates for entropy production contributions are, in principle, inconsistent. If the purely "adiabatic" numerical error in the entropy budget discussed by Johnson[4] is negligible (as in this case), the 2nd law of thermodynamics imposes energy and entropy biases, which are approximately related as $\Delta_E \approx \langle\Theta\rangle\Delta_S$. Under these hypotheses, as thoroughly discussed[18], we find that the two thermodynamic temperatures $\Theta^{+}$ and $\Theta^{-}$ are still well defined, as the expression $\left(\Theta^{+} - \Theta^{-}\right)/\Theta^{+}$ provides a good first order approximation to the true efficiency $\eta = \overline{\dot{W}}/\overline{\dot{\Phi}^{+}} = \left(\overline{\dot{\Phi}^{+}} + \overline{\dot{\Phi}^{-}} - \Delta_E\right)/\overline{\dot{\Phi}^{+}}$ ). Similarly, the material entropy production rate is actually computed as $\overline{\dot{S}_{in}} + \Delta_S$, and the irreversibility factor is evaluated as $\alpha = \left(\overline{\dot{S}_{in}} + \Delta_S\right)/\left(\eta\left(\overline{\dot{\Sigma}^{+}} + \left|\overline{\dot{\Sigma}^{-}}\right|\right)/2\right)$, where we introduce a correction in the denominator to account for the fact that $\overline{\dot{\Sigma}^{+}} \neq \left|\overline{\dot{\Sigma}^{-}}\right|$. Using these corrections, the formulas presented in section 2 apply with a high degree of approximation (increasing with decreasing efficiency) also in the case of a climate model featuring biases in the energy and entropy budgets.

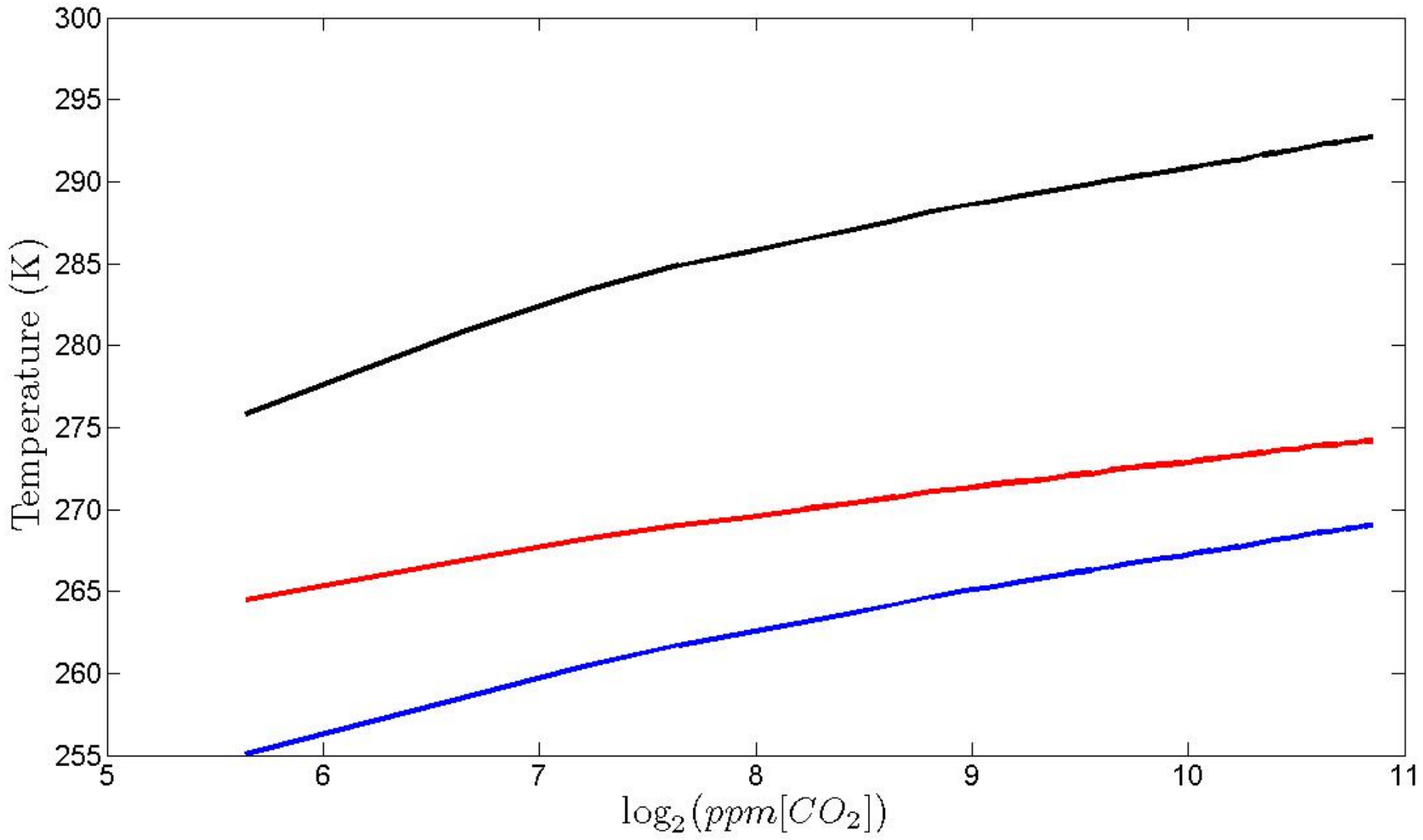

Fig. 1: Time average of the global mean surface temperature $T_s$ (black line) and of the temperature of the reservoirs at higher ($\Theta^+$) and lower ($\Theta^-$) temperature (red and blue lines, respectively).

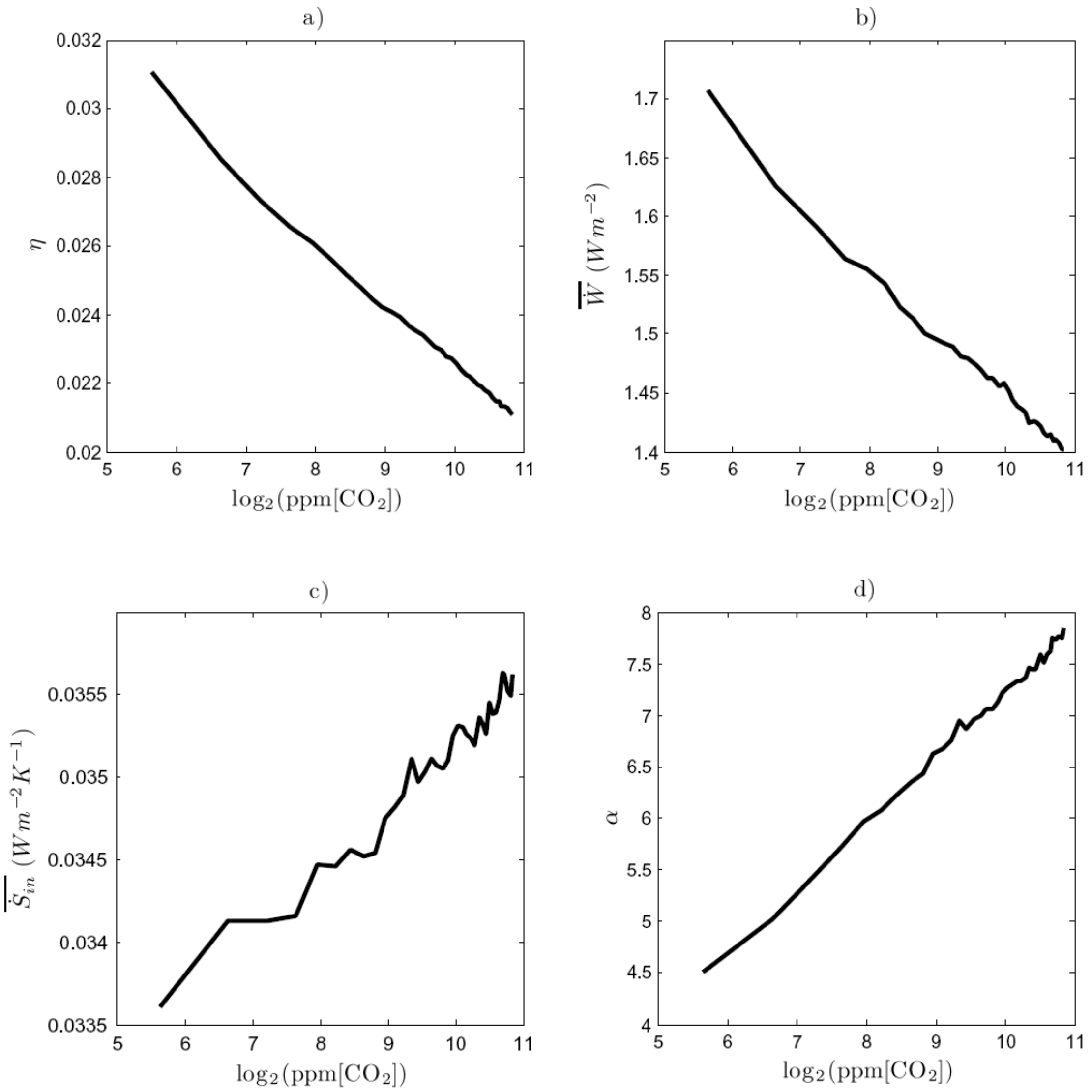

Fig. 2: Climate system sensitivities: [$CO_2$] dependence of a) thermodynamic efficiency; b) time average of the global work; c) time average of the material entropy production; d) time average of the degree of irreversibility of the system ($\alpha$ = 0 for a reversible system).

Table 1 Generalized Sensitivities

| Definition | Value |
|---|---|
| $\Lambda_{\overline{T_S}}$ | 2.55 K |
| $\Lambda_{\Theta_1}$ | 1.65 K |
| $\Lambda_{\Theta_2}$ | 2.35 K |
| $\Lambda_{\eta}$ | -0.002 |
| $\Lambda_{\overline{\dot{W}}}$ | -0.06 $Wm^{-2}$ |
| $\Lambda_{\overline{\dot{S}_{in}}}$ | 0.0004 $Wm^{-2}K^{-1}$ |
| $\Lambda_{\alpha}$ | 0.7 |

Table 2 Parameterizations

| Definition | Value |
|---|---|
| $d\Theta_1 / d\overline{T_s}$ | 0.65 |
| $d\Theta_2 / d\overline{T_s}$ | 0.92 |
| $d\eta / d\overline{T_s}$ | -0.0008 $K^{-1}$ |
| $d\overline{\dot{W}} / d\overline{T_s}$ | -0.024 $Wm^{-2}K^{-1}$ |
| $d\overline{\dot{S}_{in}} / d\overline{T_s}$ | 0.00016 $Wm^{-2}K^{-2}$ |
| $d\alpha / d\overline{T_s}$ | 0.275 $K^{-1}$ |